\documentclass[aps,prb,twocolumn,superscriptaddress,floatfix]{revtex4}
\usepackage{graphicx}
\usepackage{amsmath}
\usepackage{amssymb}
\usepackage{epsfig}
\usepackage{graphicx}
\usepackage{dcolumn}
\usepackage{bm}

\begin{document}

\title{Geometric and impurity effects on quantum rings in magnetic fields}
\author{M.~Aichinger}
\affiliation{Institut f\"ur Theoretische Physik, Johannes Kepler 
Universit\"at, A-4040 Linz, Austria}
\affiliation{Upper Austrian Research, Department for 
Medical-Informatics, Hauptstrasse 99, A-4232 Hagenberg, Austria}
\author{S.~A.~Chin}
\email[Electronic address: ]{chin@physics.tamu.edu}
\affiliation{Department of Physics, Texas A\&M University College Station,
TX 77843-4242}
\author{E.~Krotscheck}
\email[Electronic address: ]{eckhard.krotscheck@jku.at}
\affiliation{Institut f\"ur Theoretische Physik, Johannes Kepler 
Universit\"at, A-4040 Linz, Austria}
\author{E.~R\"as\"anen}
\email[Electronic address: ]{esa@physik.fu-berlin.de}
\affiliation{Institut f\"ur Theoretische Physik, Johannes Kepler 
Universit\"at, A-4040 Linz, Austria}
\affiliation{Institut f{\"u}r Theoretische Physik,
Freie Universit{\"a}t Berlin, Arnimallee 14, D-14195 Berlin, Germany}

\date{\today}

\begin{abstract}
We investigate the
effects of impurities and changing ring geometry on the
energetics of quantum rings under different magnetic field strengths.
We show that as the magnetic field and/or the electron number
are/is increased, both the quasiperiodic Aharonov-Bohm oscillations
and various magnetic phases become insensitive to 
whether the ring is circular or square in shape. 
This is in qualitative agreement with experiments. 
However, we also find that the Aharonov-Bohm oscillation can be
greatly phase-shifted by only a few impurities and can be completely
obliterated by a high level of impurity density.
In the many-electron calculations we
use a recently developed fourth-order imaginary time projection
algorithm that can exactly compute the density matrix of a free-electron
in a uniform magnetic field.
\end{abstract}

\maketitle

\section{Introduction}

The possibilities of utilizing the Aharonov-Bohm (AB) effect 
in future nanotechnology has stimulated significant recent progress
in fabricating nanoscopic quantum rings.~\cite{lorke,fuhrer,keyser} 
Since the capabilities to control
the electron number in the ring and to modify its geometry are
essential for observing and exploiting new phenomena, 
systems with these capabilities have 
attracted much theoretical interest.

Lorke and co-workers~\cite{lorke} have applied self-assembly
techniques to create InGaAs and GaAlAs/GaAs rings containing only a
few electrons. The weak electron-electron interaction in these rings
makes them most suitable for optical experiments, and the observed
state transitions can be well explained with the single-electron
spectrum of a parabolic ring.~\cite{pekka} On the other hand, Keyser
and co-workers~\cite{keyser} have reached the strongly correlated regime
by omitting the screening gate on top of a few-electron quantum ring
fabricated from a GaAs/AlGaAs heterostructure. They were able to
observe fractional AB oscillations with a period of $\Phi_0/N$, where
$\Phi_0=h/e$ is the flux quantum and $N$ is the electron number.
Exact diagonalization of a few-electron Hamiltonian~\cite{niemela}
has shown that the electron-electron interactions break the
degeneracy between the singlet and triplet states, leading to
fractional oscillations. Similar results have been obtained 
within the Heisenberg model~\cite{deo} and also by recent
Monte Carlo calculations.~\cite{emperadornew} Both have clarified the
role of the electron localization in the fractional AB effect. The
role of the (two-dimensional) width  of the ring is, however, still
unclear in the strong-interaction limit.

From magnetotransport experiments in the Coulomb blockade regime
one can infer the discrete energy levels of a quantum
ring.~\cite{fuhrer} Moreover, these measurements have been performed
for both circular and square ring-geometries; the latter corresponds
to a chaotic Sinai billiard,~\cite{kaaos} i.e., a circular
antidot at the center of a square quantum dot. In such a
symmetry-broken quantum ring, the increasing magnetic field induces
regularity in the amplitude and position of the Coulomb peaks. These
dots have been estimated to contain hundreds of electrons. The
interactions were screened by a top gate, hence the simple
single-electron picture provides a sufficient description of the
spectrum.~\cite{fuhrer}

We examine in this paper the energetics of circular, square-shaped,
and impurity-doped
two-dimensional (2D) quantum rings containing up
to $N\sim 20$ strongly interacting electrons. 
We shall focus on the 
measurable quantities such as the chemical
potentials, addition energies, and the magnetization. We will identify
the quasiperiodic AB oscillations as well as different magnetic
phases, and find, in agreement with the experiments,~\cite{fuhrer}
that these become very similar between circular and square rings at
large $N$ and in high magnetic fields. In these phases we find
similarities to integer and fractional quantum Hall states of quantum
dots.~\cite{oosterkamp,henri} We also carry out a statistical
analysis of the addition-energies for quantum rings containing
randomly distributed Coulombic impurities. Increasing the number of
impurities leads to a systematic phase shift in the AB
oscillations. As a result of the electron-electron interactions, the
high-disorder limit is characterized by a Gaussian-like
addition-energy distribution. This shape indicates 
the disappearance of the AB oscillations.

In this work, we solve the many-electron problem in a
strong, uniform magnetic field by use of the spin-density-functional
theory (SDFT). 
To solve the Kohn-Sham (KS) equations for thousands of 
impurity
configurations, we use our recently developed fourth-order
projection algorithm~\cite{magalg} to determine the occupied 
KS orbitals. This algorithm is highly efficient, since the
number of Fast Fourier Transforms used for solving the KS
spectrum remains the same even in the presence of a 
magnetic field. This is because at its core, our algorithm is
capable of exactly computing the density matrix of a free electron
in an arbitrarily strong magnetic field. Thus unlike other methods
of solving the Schr\"odinger equation, the magnetic field part of 
the physics is hardwired into our algorithm. Also, instead of the usual 
slow-convergent, charge-mixing iterations, we update the charge densities
by our linear-response algorithm\cite{cluster} with accelerated 
Newton-Raphson convergence. These significant algorithmic
advances are not based incremental improvement of numerical methods, 
but in attuning to the fundamental physics of the problem. 
The details of the algorithms are presented in
the Appendix. 

\section{Quantum-ring model} \label{model}

\begin{figure}
\includegraphics[width=8cm]{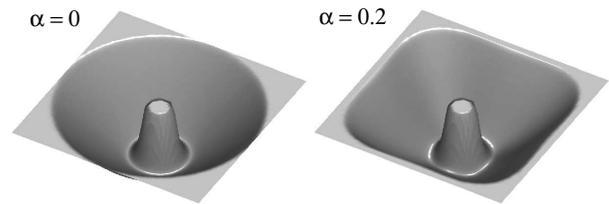}
\caption{Shapes of the external confinement potentials determining circular 
and square quantum rings with $\alpha=0$ and $0.2$ in 
Eq.~\ref{potential}, respectively.}
\label{pots}
\end{figure}

We focus on quantum rings realized in semiconductor
heterostructures, 
which can be modeled by localizing 
electrons to a 2D ({\em xy}) plane.  We use the
effective-mass approximation with GaAs parameters, i.e., the
effective electron mass $m^*=0.067\,m_e$ and the dielectric constant
$\kappa=12.7$.  The many-electron Hamiltonian is written in SI units
as
\begin{eqnarray}
H & = & \frac{1}{2m^*}\sum^N_{i=1}\left[-i\hbar\nabla_i+e\mathbf{A}
({\mathbf r}_i)\right]^2
+\sum^N_{i<j}\frac{e^2}
{4\pi\epsilon_0\kappa|{\mathbf r}_i-{\mathbf r}_j|} \nonumber \\
& + & \sum^N_{i=1}\left[V_{\rm ext}(r_i,\theta_i)+
V_{\rm imp}({\mathbf r}_i)+g^*\mu_BBs_{z,i}\right].
\label{hami}
\end{eqnarray}
The magnetic field ${\mathbf B}=B\hat{z}$ is chosen perpendicular to
the {\em xy} plane, the vector potential is then, in linear gauge,
$\mathbf{A} = -By\,\mathbf{e}_x$. The last term is the Zeeman energy
that couples the external magnetic field with the electron spin. Here
$g^*=-0.44$ is the effective gyromagnetic ratio, $\mu_B=e\hbar/2m_e$
is the Bohr magneton, and $s_z=\pm\frac{1}{2}$ for the up and down
spins, respectively.  The spin-orbit interaction is expected to be
negligible in the GaAs structure having a wide band gap, it is
therefore ignored in the Hamiltonian.  The external potential that
confines the electrons and defines the geometry of the quantum ring is
chosen, in polar coordinates, to be
\begin{equation}
V_{\rm ext}(r,\theta)=\frac{1}{2}m^*\omega_0^2
r^2\left[1+\alpha\cos(p\,\theta)\right]+V_0 e^{-r^2/d^2},
\label{potential}
\end{equation}
where $\hbar\omega_0=5$ meV is the confinement strength, and the
cosine term defines the confinement geometry. We apply a circular and
square shape determined by $\alpha=0$ and $(\alpha,p)=(0.2,4)$,
respectively.~\cite{serra} The Gaussian term in Eq.~\ref{potential}
defines an antidot at the center, thus producing a ring-like
shape of our system. We set $V_0=200$ meV and the width parameter
$d=10$ nm, which is a sufficient value for two-dimensionality.
The shapes of the total external potentials for the both
geometries are shown in Fig.~\ref{pots}.

Within the circular ring geometry ($\alpha=0$), we also apply an
impurity potential describing repulsive Coulombic impurities
located randomly in the vicinity of the quantum-ring system.
It is written as
\begin{equation}
V_{\rm imp}({\mathbf r})=\sum^{N_{\rm imp}}_{k=1}\frac{-e}
{4\pi\epsilon_0\kappa\sqrt{({\mathbf r}-{\mathbf R}_k)^2+d^2_k}},
\label{imppot}
\end{equation}
where $N_{\rm imp}$ is the number of impurities,
and ${\mathbf R}_k$ and $d_k$ are their random
lateral and vertical positions in the ranges of
$0\leq R_k\leq 100\,{\rm nm}$ and $0\leq d_k\leq 10\,{\rm nm}$, respectively.
This model, motivated by single-electron tunneling 
experiments,~\cite{jens} has been applied 
recently in statistical studies on quantum dots.~\cite{michael1,michael2}
Similarly to those studies, for each $N_{\rm imp}=5\ldots 30$
we apply $1000$ spatial configurations in order to obtain good statistics.

\section{Single-electron spectra} \label{singlespec}

\begin{figure}
\includegraphics[width=6cm]{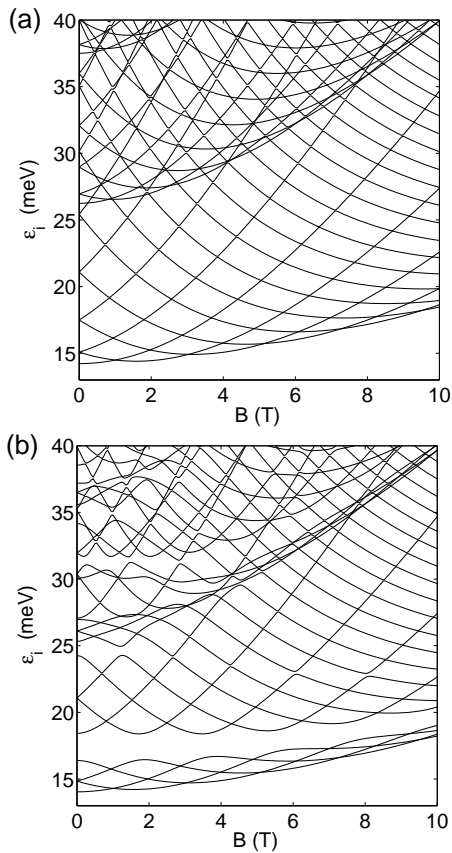}
\caption{Single-electron spectra in
circular (a) and square (b) quantum rings as a function of the
magnetic field. The Zeeman splitting is omitted.}
\label{single}
\end{figure}

In order to obtain insight into the energy-level structure in the
quantum rings studied, we first computed the single-electron spectra
of a system of noninteracting electrons as a function of the
magnetic-field strength. Figure~\ref{single}
shows the eigenenergies $\epsilon_i$ for a circular (a) and square (b)
quantum ring with up to $i\sim 25$ and $B=10$ T.  The spectrum of our
circular ring is similar to the one obtained using a pure parabolic
2D ring model.~\cite{pekka} Due to the finite ring width,
the second Fock-Darwin level and the beginning of the third one can be
seen in the spectrum. Increasing the width by making $d$ smaller
brings the upper Landau levels lower in energy.

In the spectrum of a square ring shown in Fig.~\ref{single}(b), the
four lowest energy levels are decoupled from the upper levels and form
a braid-like structure as a function of the magnetic field. Similar
decoupling can be observed for the next four levels at low fields.
This behavior results from the four-fold symmetry of the square
confining potential. 
The probability densities of the lowest eigenstates show that the four 
lowest levels correspond to energetically stable corner
modes. The next four levels in the lower row, instead, correspond to
the side modes.  Val\'{\i}n-Rodr\'{\i}quez and co-workers~\cite{serra} have
analyzed modes of this type in their far-infrared-absorption studies
on triangular and square quantum dots.


In addition to the decoupled single-electron levels, there are several
other avoided level crossings in the spectrum of a square ring,
particularly at low magnetic fields. The level repulsion in quantum
dots is usually interpreted as a signature of quantum
chaos.~\cite{kaaos} Contrary to the square quantum
dot,~\cite{esaphysica} our square-ring system is non-integrable also
at zero magnetic field. A similar system having steep walls (Sinai
billiard) is a famous example of a chaotic system, it has been
extensively studied in the context of both classical and quantum
billiards.~\cite{kaaos} In Fig.~\ref{single}(b) it can be seen that
the number of avoided crossings decreases at high magnetic fields
because the electron orbitals shrink due to the increasing magnetic
confinement. However, the system remains chaotic, again in contrast
with a square quantum dot which becomes integrable at
$B\rightarrow\infty$.

\begin{figure}
\includegraphics[width=7cm]{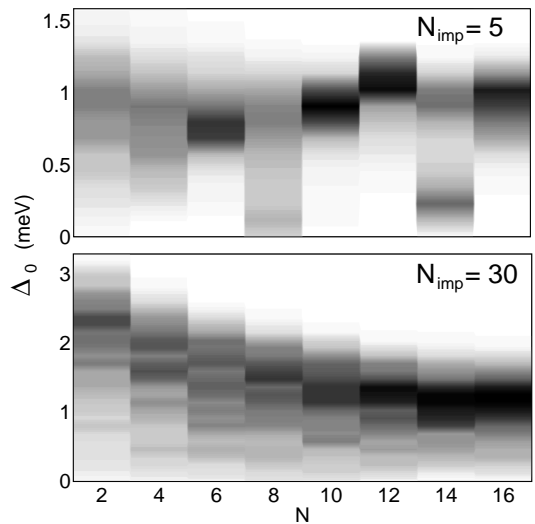}
\caption{Level-spacing distributions of noninteracting
electrons  
at $B=10$ T in impurity-affected quantum rings. 
The level spacings are calculated from one
thousand random impurity configurations for $5$ (upper panel) 
and $30$ impurities (lower panel).
}
\label{singleimpur}
\end{figure}

The presence of external impurities (see Eq.~\ref{imppot})
lead to irregular deviations in the 
single-electron energy levels. In Fig.~\ref{singleimpur} 
we plot distributions (gray scale) of single-electron level 
spacings at $B=10$ T calculated from one thousand random 
impurity configurations for $N_{\rm imp}=5$ and $30$, respectively. 
The level spacings correspond to the
addition energies of noninteracting electrons defined as 
$\Delta_0(N)=\epsilon_{N/2+1}-\epsilon_{N/2}$.
In the case of five impurities the distributions are 
approximately peaked around the level spacings of the
clean quantum ring (cf. Fig.~\ref{single} at $B=10$ T).
However, when $N=14$ ($\epsilon_8-\epsilon_7$), for example,
there are two maxima in the distribution as seen in the upper 
panel of Fig.~\ref{singleimpur}.
These result from 
the fact that while these two states are nearly degenerate
at that field, they have a different slope with respect to $B$. 
Hence, in strongly disordered configurations the level splitting 
is considerably larger than in the approximately clean cases,
leading to a two-peak structure. 
When the number of impurities is increased to 30 (lower panel
in Fig.~\ref{single}), the signatures of the shell structure clearly 
disappear. On a coarse scale, the distributions resemble
Wigner-Dyson forms indicating a disordered system.
However, the distributions show fine structure resulting
from the correlation between consecutive energy levels 
in the presence of a high number of impurities.

\section{Many-electron properties} \label{secchem}

\subsection{Circular and square geometries}

\begin{figure}
\includegraphics[width=7.5cm]{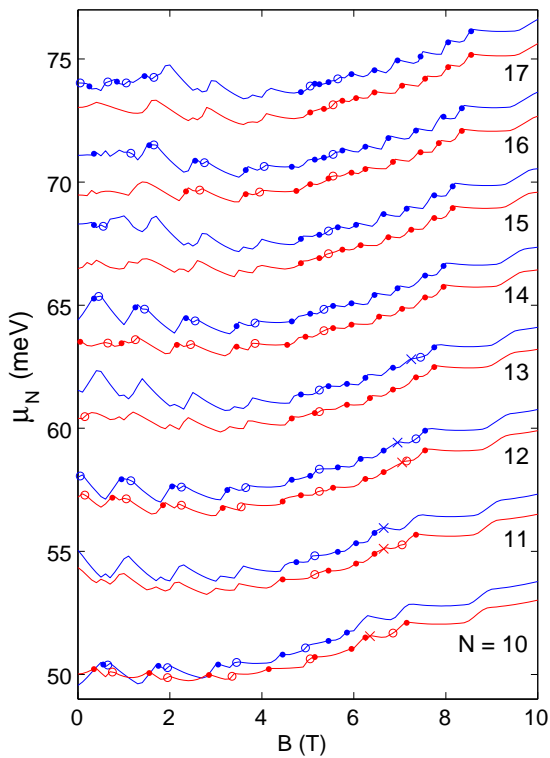}
\caption{(color online). Chemical potentials for $N=10\ldots 17$ in 
circular (upper blue lines) and square (lower red lines) 
quantum rings. The filled circles, open circles, and crosses 
correspond to the spin changes $S\rightarrow S+1$, $S\rightarrow S-1$,
and $S\rightarrow S+2$, respectively.}
\label{chem}
\end{figure}

For the corresponding many-electron problem we apply the SDFT.
Using the numerical scheme outlined in the Appendix, 
we have calculated the total energies of different spin states for
$N=1\ldots 17$ and for each magnetic-field strength up to $10$ T in
steps of $0.2$ T. The ground state for each $N$ is then defined as the
spin state having the lowest energy $E(N)$. 
Figure~\ref{chem}
shows the chemical potentials $\mu(N)=E(N)-E(N-1)$ for $N=10\ldots 17$
in circular (upper blue lines) and square (lower red lines) quantum
rings. We omit the low electron numbers in Fig.~\ref{chem} to display the 
differences between two rings more clearly. 
The transitions in the ground-state spins are marked in
Fig.~\ref{chem} such that the filled circles denote an increase
$S\rightarrow S+1$, and the open circles mark a decrease $S\rightarrow
S-1$. The crosses mark an increase of $S\rightarrow S+2$. The
rightmost points correspond to full spin polarization. The chemical
potentials for $N=10, 11$ and $12$ in circular ring agree with those of
Emperador and co-workers.~\cite{emperador} The trend of pairing
between peaks for consecutive values of $N$, as well as quasiperiodic
oscillations analyzed in Ref.~\onlinecite{emperador}, are found to
continue up to higher electron numbers.  We see the predicted
violations in the pairing, e.g., at $B\sim 1$ and $2.2$ T when
$N=11-12$. These are explained by Hund's rule, causing partial spin
polarization near the level crossings in the corresponding
single-electron spectrum (see Fig.~\ref{single}).

As seen in Fig.~\ref{chem}, the evolution of $\mu(N)$ as a function of
$B$ is qualitatively similar in circular and square rings, although
the oscillations at low fields are considerably stronger in the
circular case. This difference is due to the level repulsion of the square-shaped
ring that leads to smoother behavior of
the total energies. The tendency of the symmetry-breaking to even out
the oscillations has been detected also in quantum dots containing
external impurities.~\cite{jens,gycly} The square ring shows also 
increased stability as a small increase in $\mu(N)$ when $N=4$ or $8$ 
[not plotted in Fig.~\ref{chem}]. This effect is presented and discussed 
in detail in connection with the addition energies below.

The chemical potentials could
be directly compared to Coulomb blockade oscillations of transport
measurements, but we are not aware of such experiments for quantum
rings. In the quantum dots, however, a remarkably good
agreement has been obtained between the experimental result of the
Coulomb blockade oscillations~\cite{oosterkamp} and SDFT 
calculations.~\cite{henri}

\begin{figure}
\includegraphics[width=7cm]{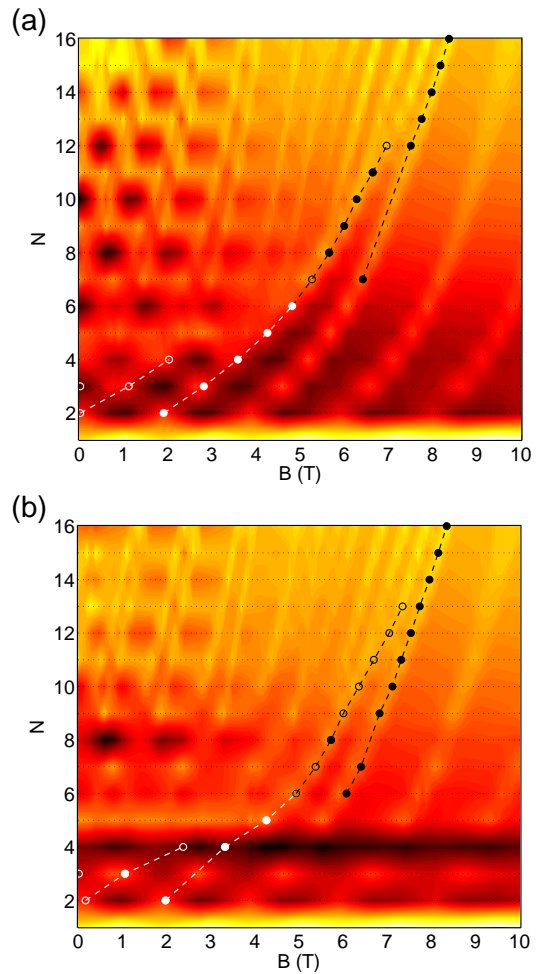}
\caption{(color online). Addition energies (yellow: low, black: high) as a
function of the magnetic field and up to 16 electrons in circular (a)
and square (b) quantum rings. The open and filled circles mark the
first and final full spin polarizations, respectively. The dashed
lines are plotted to guide the eye.}
\label{addis}
\end{figure}

Next we consider the second energy differences, i.e., the addition
energies defined as $\Delta(N)=\mu(N+1)-\mu(N)=E(N-1)-2E(N)+E(N+1)$.
Besides chemical potentials, they are also measurable
quantities,~\cite{fuhrer} and give a more detailed view on the
energetic structure of quantum rings.  Figure~\ref{addis}
shows the full phase diagram of $\Delta(N)$ as a function of $B$ and
$N$ for a circular (a) and square (b) ring. The light and dark regions
correspond to low and high addition energies, respectively. The open
and filled circles mark the first and final full spin polarizations,
and the dashed lines are to guide the eye. The characteristic energy
oscillations are evident in both geometries. The structure of the
oscillations, however, shows interesting differences.  In the circular
case there are three easily distinguished phases: (i) strong
oscillations at low $B$ that mostly correspond to spin changes
$0\rightarrow 1\rightarrow 0$ following Hund's rule; (ii) spin flip
region at intermediate $B$ that becomes broader as $N$ increases; and
(iii) a polarized regime in high magnetic fields on the right side of
the dashed line(s), characterized by large-width oscillations 
of a period $\Delta\Phi\sim \Phi_0$ (the flux $\Phi=BA$, where $A$ is
the estimated ring area).
A similar division can be found in the addition energies of
a square ring [Fig.~\ref{addis}(b)]. However, the effect of the
four-fold asymmetry is clear in the pronounced stability of the
four-electron ring. It results from the decoupling of the lowest
single-electron levels shown in Fig.~\ref{single}. The stability
is strongest after the polarization at $B\sim 3\ldots 4$ T, since then
all the four separated levels are (singly) occupied. Likewise, the
eight-electron ring is particularly stable at low fields when $S=0$.

The addition energies in high magnetic fields are very similar between the
circular and square rings, particularly when $N$ is large.
This result agrees qualitatively with the experimental observation
of Fuhrer and co-workers.~\cite{fuhrer} It can be explained 
by the fact that the magnetic confinement has a parabolic shape. 
Furthermore, the electron-electron interactions 
make the square ring effectively more symmetric.~\cite{jens}
In the onset of the full spin polarization there are two differences. First,
the square ring has a kink at $N=4$ due to its specific energy 
spectrum discussed in Sec.~\ref{singlespec}. Secondly, there are many 
depolarizations in the square ring when $N=6\ldots 13$, 
occurring between the first and second full spin polarizations
marked as open and solid circles in Fig.~\ref{addis}, 
respectively. However, this effect is relatively faint and its emergence 
is very sensitive to the ring parameters. For example, 
depolarization can be obtained in the circular ring at
$N=8\ldots 11$ if the Zeeman energy in Eq.~\ref{hami} is 
slightly reduced.



\begin{figure}
\includegraphics[width=8cm]{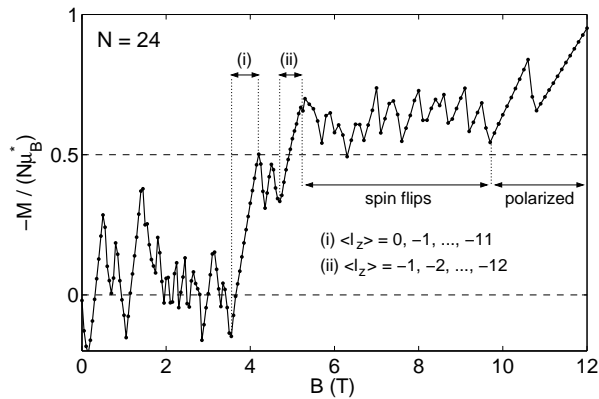}
\caption{Magnetization of a 24-electron circular quantum
ring. Different magnetic phases characteristic of
finite Fermion systems are marked in the figure.}
\label{mag24}
\end{figure}

The different magnetic phases discussed above
can be characterized in detail from the magnetization
curves for large electron numbers.
The magnetization is defined as the derivative of the free energy
with respect to the magnetic field and reduces in zero temperature
to $M=-\partial{E_{\rm tot}}/\partial{B}$. 
In Fig.~\ref{mag24}
we plot the magnetization of a 24-electron circular
quantum ring as a function of the magnetic field strength $B$.
In the two steps marked in the figure as (i) and (ii), 
the ground-state spin is zero and the (doubly) filled
angular-momentum states 
are $\left<l_z\right>=0,-1,\ldots,-11$
and $-1,-2,\ldots,-12$, respectively.
They are calculated as the expectation values of the 
angular momentum operator 
$\hat{l}_z=-i\hbar[x(\partial/\partial{y})-y(\partial
/\partial{x})]$ for different KS states.
The first occupation (i) from $\left<l_z\right>=0$ to $-(N/2-1)$
directly corresponds to a quantum Hall state~\cite{qhkirja} with a
filling factor $\nu=2$.
In the spin-flip region at $B\sim 5\ldots 10$ T 
the spin polarization of the ring
gradually increases and the magnetization 
is characterized by short-period oscillations.
In the first fully polarized $(S=12)$ state the 
occupation is $\left<l_z\right>=-2,-3,\ldots,-25$.
After that the oscillations become regular such that
each step corresponds to an increase of 24 in the 
total angular momentum $|L_z|=\sum_i{|\left<l^i_z\right>}|$,
as the electrons jump from $\left<l^i_z\right>$ to $\left<l^i_z\right>-1$.
Consequently, the hole in the electron density at the center of the 
ring increases, and there are no signs of edge reconstruction, 
which confirms the result of Emperador and co-workers.~\cite{empr}

The behavior in the electron occupations and magnetization as a
function of $B$ is similar to that of a quantum-dot
system.~\cite{henri} The most obvious difference is that the $\nu=1$
state with an occupation from $\left<l_z\right>=0$ to $-(N-1)$ is not
found in quantum rings.  This state, i.e., the maximum-density
droplet,~\cite{macdonald} is particularly stable in quantum dots and
has been observed experimentally.~\cite{oosterkamp} However, we expect
that in quantum rings the signatures of the $\nu=2$ state and spin
flips, as well as the regular oscillations in the polarized ($\nu<1$)
regime could be observed in magnetization experiments using, for
example, sensitive micromechanical magnetometers.~\cite{schwarz}


\subsection{Impurity effects}

\begin{figure}
\includegraphics[width=8cm]{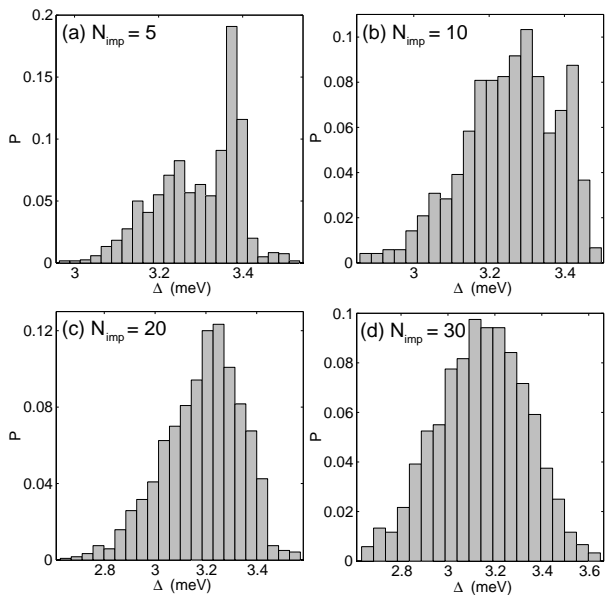}
\caption{Addition-energy distributions of a 12-electron
quantum ring having 5 (a), 10 (b), 20 (c), and 30 (d)
impurities. The magnetic field is set to $B=10$ T
corresponding to the fully-polarized regime.}
\label{distr}
\end{figure}

Finally we analyze the effect of impurities on the addition energies
in a many-electron system. The applied impurity model is
defined in Eq.~\ref{imppot}.
For simplicity, we focus here on the fully
spin-polarized regime where the addition-energy oscillations in
corresponding impurity-free quantum rings (see Fig.~\ref{addis}) are
regular. Figure~\ref{distr}
shows the addition-energy distributions of a 12-electron quantum ring
at $B=10$ T. For each number of impurities we have calculated the
electronic structure and energetics for one thousand random impurity
configurations. We find that the distribution for $N_{\rm imp}=5$
shows a clear two-peak structure. This is due to the fact that in this
case the rings can be roughly divided into relatively disordered and
clean ones depending on the actual location of the impurities.  These
two types of rings have then different radii on the average, which
eventually leads to a phase shift in the AB oscillations as a function
of the magnetic field.

As the number of impurities is increased, the two peaks gradually
merge and finally at $N_{\rm imp}=30$, corresponding to an impurity
density of $\sim 10^{-3}{\rm nm}^{-1}$, we find a Gaussian-like
symmetric distribution shown in Fig.~\ref{distr}(d).  
The symmetric shape is a result of the electron-electron interactions
that even out the tilted shape and irregularities found in the 
level-spacing distributions (noninteracting electrons)
shown in Fig.~\ref{singleimpur}. The Gaussian shape is
qualitatively similar to what has been found in
disordered quantum dots both experimentally~\cite{sivan,patel} and
theoretically using Hartree-Fock~\cite{cohen} and density-functional
methods.~\cite{hirose,jiang,michael2} In addition, modifications
to the random matrix theory that take the interactions into account
have led to a good agreement between the experiments and the 
theory,~\cite{alhassid} whereas the bare random matrix theory 
yields the Wigner-dyson distribution. Hence, 
we expect that the distribution shown in Fig.~\ref{distr}(d) for
quantum rings corresponds to the limit of disorder which ultimately 
indicates the disappearance of the AB oscillations. 
Following the analogy to quantum-dot systems,~\cite{jiang} we 
expect that a Gaussian-like distribution could be obtained in 
conductance experiments for large ($N>100$) quantum rings.
In that case the addition-energy statistics would be 
determined as a function of $N$.

\section{Summary} \label{summary}

We have used efficient computational algorithms for solving the
Kohn-Sham equations of spin-density-functional theory to examine the
effects of geometric deviations on the ground-state properties of
two-dimensional quantum rings in magnetic fields.  Due to the circular
shape of the magnetic confinement, increasing magnetic field evens out
the variations between the energetics of quantum rings of different
geometries.  The measurable quantities in both systems can be
characterized by quasiperiodic oscillations, of which we have been
able to identify the quantum-ring counterparts of the integer
quantum-Hall states.  However, the Aharonov-Bohm oscillations at high
fields are very sensitive to the presence of external impurities which
may produce a systematic phase shift. In the
high-impurity (disorder) limit the
electron-electron interactions make the addition-energy distributions
Gaussian-like. This behavior, indicating the disappearance of the
oscillations, is similar to what has been found in quantum dots.

\appendix*

\section{Numerical scheme}

Our numerical procedure for solving the electronic properties of the
system within the SDFT consists of two parts, namely, the solution of
an effective Schr\"odinger equation [the Kohn-Sham (KS) equation] and the
self-consistent determination of the spin-densities.  The KS equation
in a magnetic field is
\begin{equation}
\left[\frac{1}{2m^\star}(-i\hbar\nabla+e\mathbf{A}(\mathbf{r}))^2
+V_{\rm KS}({\mathbf r})\right]
\Phi_i^\sigma(\mathbf{r})=\epsilon_i\Phi_i^\sigma(\mathbf{r}),
\label{eq:kse}
\end{equation}
where the KS potential $V_{\rm KS}(\mathbf{r})$ is a sum of the
external potential defined above, the Hartree potential, and the
exchange-correlation potential given as $V_{\rm xc}({\mathbf
r})=\delta{E_{\rm xc}}[\rho,\xi]/\delta\rho^\sigma{({\mathbf r})}$.
Here $\rho^\sigma$ are the electron spin densities, $\sigma$ denotes
the spin index, and $\xi(\mathbf{r})=[\rho^\uparrow(\mathbf{r})
-\rho^\downarrow(\mathbf{r})]/\rho(\mathbf{r})$ is the local spin
polarization. For $E_{\rm xc}$ we use the local spin-density
approximation with the functional provided by Attaccalite and 
co-workers.~\cite{AMGB02}

The lowest $n$ solutions of the eigenvalue problem (\ref{eq:kse}) are
obtained by applying the evolution operator,
\begin{equation}
\mathcal{T}(\epsilon)\equiv\text{e}^{-\epsilon H}
\end{equation}
repeatedly to a set of states $\{\psi_j,\, 1\leq j\leq n\}$, and
orthogonalizing the states after every step.  Instead of the commonly
used second-order factorization in combination with the Gram-Schmidt
orthogonalization, we use the fourth-order factorization for the
evolution operator given by~\cite{Suzuki96,ChinPLA97}
\begin{equation}
\mathcal{T}^{(4)}(\epsilon)=e^{-\frac{1}{6}\epsilon V}
e^{-\frac{1}{2}\epsilon T}e^{-\frac{2}{3}\epsilon\widetilde V}
e^{-\frac{1}{2}\epsilon T}e^{-\frac{1}{6}\epsilon V}+
\mathcal{O}(\epsilon^5)\,,
\label{eq:4ord}
\end{equation}
where
\begin{equation}
T = \frac {1}{2 m^\star}\mathbf{\Pi}^2 \equiv \frac {1}{2 m^\star}
\left[\Pi_x^2 + \Pi_y^2\right], \quad  \mathbf{\Pi} = {\bf p} + e{\bf A}(\mathbf{r})
\end{equation}
is the kinetic-energy operator. We have
defined the local
modified
potential~\cite{Suzuki96,ChinPLA97} as
\begin{equation}
\widetilde V=V+\frac{1}{48}\epsilon^2[V,[T,V]].
\label{eq:so4}
\end{equation}
Note that the vector potential ${\bf A}(\mathbf{r})$ does
not contribute to the commutator.

We have shown in Ref.~\onlinecite{magalg} that 
the density matrix
$e^{-\epsilon T}$ 
can be exactly decomposed for a uniform
magnetic field as
\begin{eqnarray}
e^{-\frac{\epsilon}{2 m^\star}(\Pi_x^2 + \Pi_y^2)} & = &
e^{-\frac{\epsilon}{2 m^\star} C_E(\xi) \Pi_x^2}
e^{-\frac{\epsilon}{2 m^\star} C_M(\xi)\Pi_y^2}\times\nonumber\\
& \times &
e^{-\frac{\epsilon}{2 m^\star} C_E(\xi)\Pi_x^2},\label{eq:evop}
\end{eqnarray} 
where $\xi = \epsilon \hbar e B/m^*$, and
\begin{equation}
C_E(\xi)=\frac{\cosh(\xi)-1}{\xi\sinh(\xi)}
\quad
\text{and}\quad
C_M(\xi)=
\frac{\sinh(\xi)}{\xi}\,.
\end{equation}
The exact factorization shown in Eq. (\ref{eq:evop}) is possible
because the Hamiltonian of an electron in a uniform magnetic field is
quadratic, and higher-order commutators of $\Pi_x^2$ and $\Pi_y^2$ are
either zero or simply proportional to $\Pi_x^2$ and $\Pi_y^2$. The two
key commutators are
\begin{equation}
[\Pi_i^2,[\Pi_j^2,\Pi_i^2]]=8\hbar^2 e^2 B^2\Pi_i^2.
\label{eq:hotv}
\end{equation}
Hence, all higher order commutators appearing in the
Baker--Campbell--Hausdorff formula can be summed back to the original
operators $\Pi_x^2$ and $\Pi_y^2$.

The above result can be easily generalized to a charged particle in a
uniform magnetic field in an arbitrary external potential:
Inserting the exact factorization (\ref{eq:evop}) into the
factorization (\ref{eq:4ord}) of the full Hamiltonian
yields the final result:
\begin{eqnarray}
\mathcal{T}^{(4)}(\epsilon) & = & e^{-\frac{1}{6}\epsilon V}
e^{-\frac{\epsilon}{4 m^\star}C_E(\frac{\xi}{2})\Pi_x^2}
e^{-\frac{\epsilon}{4 m^\star}C_M(\frac{\xi}{2})\Pi_y^2}\times\nonumber\\
& \times &
e^{-\frac{\epsilon}{4 m^\star}C_E(\frac{\xi}{2})\Pi_x^2}
e^{-\frac{2}{3}\epsilon\widetilde V}
e^{-\frac{\epsilon}{4 m^\star}C_E(\frac{\xi}{2})\Pi_x^2}\times \\
& \times &
e^{-\frac{\epsilon}{4 m^\star}C_M(\frac{\xi}{2})\Pi_y^2}
e^{-\frac{\epsilon}{4 m^\star}C_E(\frac{\xi}{2})\Pi_x^2}
e^{-\frac{1}{6}\epsilon V}+
\mathcal{O}(\epsilon^5)\,.\nonumber \label{eq:4ex}
\end{eqnarray}
In Ref.~\onlinecite{magalg} we have shown that this algorithm
is, depending on the system and the desired accuracy, a factor
of 10 to 100 more efficient than the second order factorization.
There, we have worked in circular gauge, here we point out that
the method becomes even more efficient in linear gauge.
The algorithm is then applied as follows:
\begin{itemize}
\setlength{\labelwidth}{5truecm}
\item[(1)] Start with a set of suitably-chosen 
initial states in the coordinate space.
\item[(2)] Multiply these states by
$e^{-\frac{1}{6}\epsilon V}$.
\item[(3)] Fourier transform the $x$-coordinate of each state
to the $k_x$-space and
multiply by  $e^{-\frac{\epsilon}{4 m^\star}
C_E(\frac{\xi}{2})(k_x+eBy)^2}$.
\item[(4)]
Fourier transform now
$y$ to the $k_y$-space, and
multiply by $e^{-\frac{\epsilon}{4 m^\star}
C_M(\frac{\xi}{2})k_y^2}$.
\item[(5)] Do the inverse transformation back to $y$
and multiply by  $e^{-\frac{\epsilon}{4 m^\star}
C_E(\frac{\xi}{2})(k_x+\frac{1}{2}By)^2}$.
\item[(6)] Fourier transform $k_x$ back to the $x$-space
and multiply by
$e^{-\frac{2}{3}\epsilon \tilde V}$. Then repeat the steps
(3)-(5) and finally multiply the states by
$e^{-\frac{1}{6}\epsilon V}$.
\item[(7)] Orthonormalize the states and repeat the procedure
until convergence has been obtained.
\end{itemize}
Thus, the implementation of the algorithm requires the equivalent of 
two 2D Fourier transforms. In other words it is computationally
no more costly than the case without a magnetic field.
For the orthonormalization, we diagonalize the matrix of the overlap
integrals and from these we construct a new set of orthonormal states.

A 
persistent
problem of density-functional calculations is that the
na\"ive charge density mixing iteration scheme,
%
usually require a large number of iterations for convergence.
We overcome this problem by using a method \cite{cluster} which solves 
for $\rho(\mathbf{r})$ directly by applying a Newton-Raphson procedure
is used. We define
\begin{equation}
\Delta\rho^\sigma({\bf r})=
\sum_\mathbf{h}n_\sigma(\mathbf{h})
\left|\Phi^{\star\,\sigma}_{\mathbf{h}}[\rho_\uparrow,\rho_\downarrow]
(\mathbf{r})\right|^2
-\rho^\sigma(\mathbf{r})
\end{equation}
as the density difference between two self-consistent iterations.
Here $n^\sigma$ is the occupation factor,
$\Phi^\sigma_{\mathbf{h}}$ are the orthogonalized solutions of 
Eq. (\ref{eq:kse}), and $\rho^\sigma(\mathbf{r})$ is the density used 
for the calculation of $V_{\rm KS}$. The sum goes over all occupied 
(hole: ${\bf h}$) states. Then the density correction 
$\delta\rho^{\sigma}({\bf r})$ is determined by a linear equation
\begin{equation}
\Delta\rho^\sigma({\bf r}) = \sum_{\sigma^\prime}\int
d^dr^\prime\,\varepsilon^{\sigma,\sigma^\prime}
({\bf r},{\bf r^\prime};0)
\delta\rho^{\sigma^\prime}({\bf r}^\prime)\label{eq:resp}.
\end{equation}
Here $d$ is the dimension of the system and
$\varepsilon^{\sigma,\sigma^\prime}$ is the static dielectric function
of a non-uniform electron gas.~\cite{pinesnoz}  It contains the
zero-frequency Lindhard function and the particle-hole potential
$V_{p-h}^{\sigma,\sigma^\prime}({\bf r},{\bf r^\prime})= \delta V_{\rm
KS}^\sigma({\bf r})/ \delta\rho^{\sigma^\prime}({\bf r^\prime})$.  To
avoid the calculation of unoccupied (particle: ${\bf p}$) states we
seek for an approximation for the static response function that only
needs the calculation of occupied states.  For the purpose such an
algorithm, we recall that linear response theory can be derived
\cite{thouless} from an action principle for excitations of the form
$\left|\psi(t)\right\rangle = \exp(\sum_{ph}c_{ph}(t)a^\dagger_p a_h)
\left|\phi_0\right\rangle$, where $\left|\phi_0\right\rangle$ is the
ground state, and $c_{ph}(t)$ are particle--hole amplitudes. If we
assume that the particle--hole amplitudes are matrix elements of a
{\it local\/} function $\omega^\sigma({\bf r})$, i.e.,
$c_{ph}(t) = \left\langle p\right| \omega^\sigma({\bf
r},t)\left|h\right\rangle$ we end up with Feynman's theory of
collective excitations.~\cite{feyn} Using the commutator of the kinetic
part of the effective Schr\"{o}dinger equation with $\omega^\sigma$,
\begin{eqnarray}
 \frac{1}{2}\left[\Pi^2,\omega^\sigma\right]
=&\frac{1}{2}\left[\Pi\left[\Pi,\omega^\sigma\right]+
\left[\Pi,\omega^\sigma\right]\Pi\right]\nonumber\\
=&-\frac{i}{2}
\left[\Pi\left(\mathbf{\nabla}\omega^\sigma\right)+
\left(\mathbf{\nabla}\omega^\sigma\right)\Pi\right],
\end{eqnarray}
the magnetic part completely cancels out because it is local.
Thus, applying the response-algorithm in a magnetic field
also applies no computational overhead compared to the
zero-field case. In the ``collective approximation'' we can
rewrite Eq. \ref{eq:resp} as~\cite{cluster}
\begin{eqnarray}
&\left[-\frac{1}{2}\nabla\cdot\left[\rho^\sigma({\bf r})
\nabla\right] +
2\sum_{\sigma^\prime}S_F^\sigma\,\star\,V^{\sigma,\sigma^\prime}_{p-h}
\,\star\,S_F^{\sigma^\prime}\,\star\right] w^{\sigma^\prime}\nonumber\\ 
&= 2\sum_{\sigma^\prime}S_F^\sigma\,\star\, V^{\sigma,\sigma^
\prime}_{p-h}
\,\star\,\Delta\rho^{\sigma^\prime}\label{eq:coll}
\end{eqnarray}
where now
$$
\delta\rho^\sigma({\bf r})\,=\,\Delta\rho^\sigma({\bf r})-
S^\sigma_F({\bf r},{\bf r^\prime}) \,\star\,{w}^\sigma({\bf r^
\prime}),
$$
and
$$
S_F^\sigma({\bf r},{\bf r}^\prime)
=\rho^\sigma({\bf r})\delta({\bf r}-{\bf r^\prime})-
\frac{1}{2} \left|\sum_{\bf h}\Phi^{\sigma\,\star}_{\bf h}
({\bf r})\Phi^{\sigma}_{\bf h} ({\bf r^\prime})\right|^2,
\eqno(15)
$$
is the static structure function of the noninteracting system.  Above,
the asterisk stands for the convolution integral. With these
manipulations, we have rewritten the response-iteration equation in a
form that requires only the calculation of the occupied states.  Since
the multiplication on the left-hand side of Eq. \ref{eq:coll} requires only
vector--vector operations, the equation can be solved either directly
or with iterative methods like the conjugate-gradient method or
multigrid methods.~\cite{aithesis}

\begin{acknowledgments}
This work was supported by the Austrian Science Fund under project 
No. P15083-N08 (to E. K.), by the U. S. National Science Foundation
under grant No. DMS-0310580 (to S. A. C.), and by the NANOQUANTA NOE
and the Finnish Academy of Science and Letters, Vilho, Yrj{\"o} and 
Kalle V{\"a}is{\"a}l{\"a} Foundation (to E. R.).
Generous computational support
was provided by the Central Computing Services at the Johannes Kepler
Universit\"at Linz, we would especially like to thank Johann Messner
for help and advice in using the facility. E. R. thanks H. Saarikoski
for useful discussions.
\end{acknowledgments}


\begin{thebibliography}{10}

\bibitem{lorke}
A. Lorke, R.~J. Luyken, A.~O. Govorov, J.~P. Kotthaus, J.~M. Garcia, and P.~M.
  Petroff, Phys. Rev. Lett. {\bf 84},  2223  (2000).

\bibitem{fuhrer}
A. Fuhrer, S. L{\"u}scher, T. Ihn, T. Heinzel, K. Ensslin, W. Wegscheider, and
  M. Bichler, Nature (London) {\bf 413},  822  (2001).

\bibitem{keyser}
U.~F. Keyser, C. F{\"u}hner, S. Borck, R.~J. Haug, M. Bichler, G. Abstreiter,
  and W. Wegscheider, Phys. Rev. Lett. {\bf 90},  196601  (2003).

\bibitem{pekka}
T. Chakraborty and P. Pietil{\"a}inen, Phys. Rev. B {\bf 50},  8460  (1994).

\bibitem{niemela}
K. Niemel{\"a}, P. Pietil{\"a}inen, P. Hyv{\"o}nen, and T. Chakraborty,
  Europhys. Lett. {\bf 36},  533  (1996).

\bibitem{deo}
P.~S. Deo, P. Koskinen, M. Koskinen, and M. Manninen, Europhys. Lett. {\bf 63},
   846  (2003).

\bibitem{emperadornew}
A. Emperador, F. Pederiva, and E. Lipparini, Phys. Rev. B {\bf 68},  115312
  (2003).

\bibitem{kaaos}
H.-J. Stockmann, {\em Quantum Chaos: An Introduction} (Cambridge University
  Press, Cambridge, 2000).

\bibitem{oosterkamp} T. H. Oosterkamp, J. W. Janssen,
L. P. Kouwenhoven, D. G. Austing, T. Honda, and S. Tarucha,
Phys. Rev. Lett. {\bf 82}, 2931 (1999).

\bibitem{henri}
H. Saarikoski and A. Harju, Phys. Rev. Lett. {\bf 94}, 246803 (2005).

\bibitem{magalg}
M. Aichinger, S.~A. Chin, and E. Krotscheck, Comp. Phys. Comm {\bf  171}, 197 (2005).

\bibitem{cluster}
M. Aichinger and E. Krotscheck, Comp. Mater. Sci. {\bf 34}, 188 (2005).

\bibitem{serra}
M. Val\'{\i}n-Rodr\'{\i}quez, A. Puente, and L. Serra, Phys. Rev. B {\bf 64},
  205307  (2001).

\bibitem{jens} E. R{\"a}s{\"a}nen, J. K{\"o}nemann, R.~J. Haug, M.~J. Puska, and R.~M.
  Nieminen, Phys. Rev. B {\bf 70},  115308  (2004).

\bibitem{michael1}
M. Aichinger and E. R{\"a}s{\"a}nen, Phys. Rev. B {\bf 71}, 165302 (2005).

\bibitem{michael2}
E. R{\"a}s{\"a}nen and M. Aichinger, Phys. Rev. B {\bf 72}, 045352 (2005).

\bibitem{esaphysica}
See, e.g., E. R{\"a}s{\"a}nen, M. J. Puska, and R. M. Nieminen, Physica E
  (Amsterdam) {\bf 22} 490 (2004). The classical case has been studied by M.
  Robnik and M. V. Berry, J. Phys. A {\bf 18}, 11361 (1985).

\bibitem{emperador}
A. Emperador, M. Pi, M. Barranco, and E. Lipparini, Phys. Rev. B {\bf 64},
  155304  (2001).

\bibitem{gycly}
A.~D. G{\"u}\c{c}l{\"u}, J.~S. Wang, and H. Guo, Phys. Rev. B {\bf 68},  035304
   (2003).

\bibitem{qhkirja} T. Chakraborty and P. Pietil\"ainen, {\em The Quantum
Hall Effects: Fractional and Integral} (Springer, Berlin, 1995).

\bibitem{empr}
A. Emperador, M. Barranco, E. Lipparini, M. Pi, and L. Serra, Phys. Rev. B {\bf 59},
  15301  (1999).

\bibitem{macdonald} A. H. MacDonald, S. R. Eric Yang, and
M. D. Johnson, Aust. J. Phys. {\bf 46}, 345 (1993).

\bibitem{schwarz}
M.~P. Schwarz, D. Grundler, C. Heyn, D. Heitmann, D. Reuter, and A. Wieck,
  Phys. Rev. B {\bf 68},  245315  (2003).

\bibitem{sivan}
U. Sivan, R. Berkovits, Y. Aloni, O. Prus, A. Auerbach, and G. Ben-Yoseph,
  Phys. Rev. Lett. {\bf 77},  1123  (1996).

\bibitem{patel}
S. R. Patel {\em et al.}, Phys. Rev. Lett. {\bf 80}, 4522  (1998).

\bibitem{cohen}
A. Cohen, K. Richter, and R. Berkovits,
Phys. Rev. B {\bf 60}, 2536  (1999).

\bibitem{hirose}
K. Hirose, F. Zhou, and N.~S. Wingreen, Phys. Rev. B {\bf 63},  75301  (2001).

\bibitem{jiang}
H. Jiang, D. Ullmo, W. Yang, and H. U. Baranger,
Phys. Rev. B {\bf 69} 235326  (2004).

\bibitem{alhassid}
Y. Alhassid, Ph. Jacquod, and A. Wobst,
Phys. Rev. B {\bf 61}, R13357  (2000).

\bibitem{AMGB02}
C. Attaccalite, S. Moroni, P. Gori-Giorgi, and G.~B. Bachelet, Phys. Rev. Lett.
  {\bf 88},  256601  (2002).

\bibitem{Suzuki96}
M. Suzuki,  in {\em Computer Simulation Studies in Condensed Matter Physics},
  edited by D.~P. Landau, K.~K. Mon, and H.-B. Sch{\"u}ttler (Springer, Berlin,
  1996), Vol.~VIII, pp.\ 1--6.

\bibitem{ChinPLA97}
S.~A. Chin, Phys. Lett. A {\bf 226},  344  (1997).

\bibitem{pinesnoz}
D. Pines and P. Nozieres, {\it The Theory of Quantum Liquids}
(Benjamin, New York, 1966).

\bibitem{thouless}
D.~J. Thouless, {\it The Quantum Mechanics of Many-body Systems}, 2
ed, (Academic Press, New York, 1972).

\bibitem{feyn}
R.~P. Feynman, {\it Statistical Mechanics - A Set of Lectures},
Benjamin Advanced Book, Reading, MA, 1972.

\bibitem{aithesis}
M. Aichinger, {\it Spin-Density-Functional-Theory Calculations of 2D
Finite Electron Systems}, PhD thesis, Johannes Kepler Universit\"at, Linz, Austria (2005).


\end{thebibliography}
\end{document}